\documentclass[twocolumn]{aastex631}
\usepackage{eqnarray,amsmath}

\begin{document}

\title{Limits of Standard Tidal Models at Quaoar: Matching Weywot’s Orbit, Missing the Spin}

\correspondingauthor{Zsolt Regály}
\email{regaly@konkoly.hu}

\author[0000-0001-5573-8190]{Zsolt Regály}
\affiliation{HUN-REN CSFK Konkoly Observatory, MTA Centre of Excellence, Konkoly Thege M. \'ut 15-17, Budapest, 1121, Hungary}

\author[0000-0003-3780-7185]{Viktória Fröhlich}

\author[0000-0002-8722-6875]{Csaba Kiss}

\affiliation{HUN-REN CSFK Konkoly Observatory, MTA Centre of Excellence, Konkoly Thege M. \'ut 15-17, Budapest, 1121, Hungary}

\affiliation{ELTE Eötvös Loránd University, Institute of Physics and Astronomy, P\'azm\'any P\'eter S\'et\'any 1/A, Budapest, 1117, Hungary}




\begin{abstract}
Weywot, Quaoar’s small satellite, follows a nearly circular orbit at a distance of 12.9 times Quaoar’s diameter and coexists with a compact ring system. 
Nevertheless, Quaoar’s flattening of 0.16, slow 17.7~hr rotation and Weywot’s low mass are difficult to reconcile with conventional tidal-evolution theory.
We assess whether standard tides can reproduce the present‐day architecture of the Quaoar-Weywot system and identify the initial conditions required.
Orbit‑averaged integrations spanning 4.5~Gyr were carried out with two formalisms: (i) a constant phase‑lag (CPL) and (ii) an Andrade creep‑tide (ACT) framework.
With the nominal Weywot mass, both tidal prescriptions converge on Weywot’s observed orbital distance for a wide range of initial orbital distances and eccentricities; eccentricity is damped and present‑day tidal torques are negligible, rendering the orbit quasi‑stationary. 
Quaoar’s spin, however, remains essentially unchanged from its inferred primordial period based on its present-day flattening, and does not reproduce the observed value. A match is possible only if Weywot is $\gtrsim5-10\times$ more massive than current estimates and if its initial eccentricity is finely tuned; such scenarios are inconsistent with occultation‑derived masses and imply an implausibly dense satellite.
Based on the best fitting viscoelastic parameters, the most plausible composition for Quaoar is found to be a partially differentiated dwarf planet containing roughly equal masses of silicate rock and $\mathrm{H_2O}$‑dominated warm (150-180~K) ices.
Standard tidal models reproduce Weywot’s semimajor axis but cannot account for Quaoar’s slow 17.7~hr rotation without invoking an unrealistically massive satellite or external torques, suggesting that non-tidal processes -- such as a largely primordial spin, early satellite loss, or a retrograde secondary giant impact -- must have influenced Quaoar’s rotational evolution.

\end{abstract}



\section{Introduction} \label{sec:intro}

The discovery of large trans-Neptunian objects (TNOs) in the outer Solar System has substantially advanced our understanding of the composition, dynamical behavior, and evolutionary history of regions beyond Neptune. 
Among these bodies, (50000) Quaoar,
a dwarf planet in the Kuiper Belt (KB)
stands out as particularly compelling \citep{Brown2007IAUC.8812....1B}.
Quaoar is characterized by a low albedo and a composition dominated by a mixture of rock and ice, distinguishing it from the more volatile-rich bodies of the inner Solar System \citep{Fraseretal2010ApJ...714.1547F}. 
By analyzing the thermal light curve, \citet{Kiss2024AA...684A..50K} revealed that Quaoar's diameter is in the range $D_\mathrm{p}=1094-1111$~km and density is in the range $\rho_\mathrm{p}=1.67-1.77~\mathrm{g~cm^{-3}}$.

Quaoar’s orbital characteristics underscore its status as a dynamically stable object within the KB. 
With a semimajor axis of approximately 43.2 AU and an orbital period of 287 years, it follows a nearly circular, low-inclination orbit -- suggesting that it has remained largely unperturbed by significant gravitational interactions throughout much of the Solar System’s history \citep{Brown2004}.

The discovery of Quaoar’s moon Weywot 
in 2007 introduced an additional dimension to the study of this distant dwarf planet \citep{Brown2007IAUC.8812....1B}. 
With an orbital period of approximately 12.4 days and a separation of about $14,060$~km from its primary, Weywot is gravitationally bound and resides within a dynamically stable configuration.
Prior to 2019, Weywot’s orbit was poorly constrained due to limited observations and minimal parallactic motion, leading to a mirror ambiguity in inclination -- making it unclear whether the orbit was prograde or retrograde. 
Sparse data also introduced a 0.39~day alias in the orbital period, allowing multiple plausible solutions. 
Earlier estimates (e.g., \citealp{Fraseretal2010ApJ...714.1547F}) suggested a high eccentricity ($e \simeq 0.14$), prompting hypotheses involving collisions, gravitational perturbations, or a collisional origin \citep{Fraser2013Icar..222..357F}. 
Some of these uncertainties were resolved after stellar occultation observations beginning in August 2019 confirmed a near-circular ($e=0.018$) orbit with a 12.4~day period \citep{Morgadoetal2023Natur.614..239M,Braga-Ribas2025RSPTA.38340200B}.
However, it remains unknown whether Weywot orbits Quaoar in a prograde or retrograde sense relative to Quaoar’s spin.
Preliminary estimates suggest that Weywot has a diameter of $\sim200$~km and a predominantly icy composition \citep{Fernandez-Valenzuelaetal2023DPS....5520204F}.

In addition to hosting the satellite Weywot, Quaoar also possesses a faint ring system \citep{Morgadoetal2023Natur.614..239M,Pereira2023A&A...673L...4P}. 
The ring system, consisting of at least two components made of small particles, is located at a distance of approximately 2,220~km and 4,074~km from the primary, well within the orbital radius of Weywot.
Recent stellar occultation measurements, however, revealed the presence of a third outer ring, or possibly an additional satellite, at a radial distance of 5,757~km, located near the 7:2 orbital resonance with Weywot \citep{Nolthenius_2025}. 
Observational analyses suggest that spin-orbit resonances involving Quaoar's rotation and mean-motion resonances with Weywot -- particularly the 6:1 commensurability -- may contribute to the confinement and long-term stability of the ring system (see, e.g., \citealp{Morgadoetal2023Natur.614..239M,Pereira2023A&A...673L...4P,Rodriguez2023MNRAS.525.3376R,Proudfoot2025PSJ.....6..146P}).

Tidal evolution models are essential for understanding the long-term dynamical stability of satellite systems around TNOs. 
Small satellites orbiting large TNOs can undergo a variety of tidal interactions, including orbital expansion or decay, spin–orbit synchronization, and, in extreme cases, inward migration that may culminate in a collision with the primary body \citep{Dones1999Icar..142..509D}.
Weywot’s present-day orbital configuration likely reflects a history of tidal interactions and may contain clues about its origin -- whether it was captured by Quaoar or formed in situ through giant impact.
Based on earlier measurements of Weywot’s eccentricity ($e\simeq0.14$) and plausible estimates of the effective tidal dissipation factor, \citet{Fraser2013Icar..222..357F} concluded that Weywot, despite having evolved into a state of synchronous rotation, likely did not acquire  orbital eccentricity through tidal evolution from an initially circular orbit. Instead, the eccentricity was either imparted at formation or subsequently excited by an external mechanism.
Note, however, that the most recent measurements suggest a very low orbital eccentricity of $e=0.018$ \citep{Morgadoetal2023Natur.614..239M,Braga-Ribas2025RSPTA.38340200B}.

In this study, we investigate the tidal evolution of the Quaoar–Weywot system over a broad range of initial orbital and rotational conditions, with the aim of explaining Quaoar’s present flattening ($f=0.16$), its slow rotation period of 17.7~hr, and Weywot’s nearly circular orbit. 
We implicitly assume that Weywot formed, possibly as a result of a giant impact, acquiring a significant initial orbital eccentricity that was subsequently damped to a near-circular state through tidal evolution over the age of the Solar System.
To this end, we employ two distinct tidal models: the constant phase-lag (CPL) and the Andrade creep-tide (ACT) framework.
Our primary objective is to determine whether the currently best-estimated orbital parameters of Weywot, along with the rotational state of Quaoar, can be reproduced under realistic and physically plausible assumptions.

The paper is organized as follows. 
Section~2 introduces the tidal models and outlines the assumptions adopted. 
A detailed description of the CPL and ACT frameworks, together with the computation of the required eccentricity functions, is provided in the Appendices. 
In Section~3, we present simulations aimed at constraining the tidal dissipation parameters for the Quaoar-Weywot system. 
Section~3 also reports a parameter study that tests the tidal models against a wide range of initial orbital and spin configurations. 
Finally, Section~4 summarizes our results and provides the main conclusions.

\section{Tidal evolution models}

The long-term tidal evolution of the Quaoar-Weywot system is modeled over a timescale of 4.5~Gyr using orbit-averaged formulations. 
Two distinct tidal frameworks are employed: (1) the method developed by \citet{Ferraz-Mello2013CeMDA.116..109F}, in combination with the constant phase lag theory of \citet{GoldreichPeale1966AJ.....71..425G} (CPL), and (2) the method of Andrade creep-tide (ACT) formulated by \citet{Boue2019CeMDA.131...30B}. 
Detailed descriptions of both approaches are provided in Appendix~\ref{sec:const-lag} and \ref{sec:nonconst-lag}.
Both tidal frameworks assume that Quaoar and Weywot are homogeneous spheres and are suitable for systems with significant orbital eccentricities. 
To ensure accuracy in such cases, we implement a method to effectively compute high-order eccentricity functions, as discussed in Appendix~\ref{apx:E2k}. 
The physical parameters adopted for the Quaoar-Weywot system throughout this study are summarized in Table~\ref{tab:physparam}.

In a circular synchronous prograde orbit\footnote{Throughout this work, the terms prograde and retrograde refer to the orbital sense relative to the spin direction of the primary.}, the satellite remains fixed relative to a single point on the primary’s equator, appearing motionless in the planet’s rotating frame.
It is defined by the condition $P_\mathrm{p} = 2\pi/n$, where $P_\mathrm{p}$ is the primary's spin period and $n$ is the satellite’s mean motion.
The synchronous orbit is given by
\begin{equation}
a_\mathrm{sync} = \left[ \left( \frac{P_\mathrm{p}}{2\pi} \right)^2 G \left(M_\mathrm{p} + M_\mathrm{s}\right) \right]^{1/3},
\end{equation}
where $M_\mathrm{p}$ and $M_\mathrm{s}$ are the masses of the primary and the satellite, respectively.
For Quaoar’s current spin period $P_\mathrm{p}=17.7$~hr, the synchronous orbit is at $a_\mathrm{sync}\simeq 2020$~km, i.e., $\simeq1.84~D_\mathrm{p}$, where $D_\mathrm{p}$ is the diameter of Quaoar.
For a more rapid rotation, for instance with $P_\mathrm{p}=5$~hr, the corresponding synchronous orbit is at $a_\mathrm{sync}\simeq 870$~km, which falls below the fluid Roche limit.

\begin{deluxetable*}{lcccccccccc}
\label{tab:physpar}
\tablecaption{Physical parameters of the Quaoar-Weywot system used for tidal calculations.\label{tab:physparam}}
\tablewidth{700pt}
\tabletypesize{\scriptsize}
\tablehead{
\colhead{\bf Object} &
\colhead{\bf Mass} &
\colhead{ {\bf Diameter$^\sharp$}} &
\colhead{\bf Density} &
\colhead{\bf Orb. dist.} &
\colhead{\bf Orb. ecc.} &
\colhead{\bf Rot. period} &
\colhead{\bf Oblateness} &
\colhead{\bf Refs.} \\
\colhead{} &
\colhead{$\mathrm{kg}$} &
\colhead{$\mathrm{km}$} &
\colhead{$\mathrm{kg~m^{-3}}$} &
\colhead{$\mathrm{km}$} &
\colhead{} &
\colhead{$\mathrm{hr}$} &
\colhead{$f$}
}
\startdata
Quaoar & $1.2\times10^{21}$ & 1090 & 1770& - & - & 17.7 & 0.16 & \citet{Morgadoetal2023Natur.614..239M}\\
&&&&&&&&\citet{Kiss2024AA...684A..50K}\\
Weywot & $4.19\times10^{18}$ & 200 & 1000 & 14,060 & 0.018 & - &  - & \citet{Fernandez-Valenzuelaetal2023DPS....5520204F}\\
&&&&&&&&\citet{Kiss2024AA...684A..50K}\\
&&&&&&&&\citet{Braga-Ribas2025RSPTA.38340200B}\\
\enddata
\tablecomments{$^\sharp$Volumetric equivalent diameter is calculated from $V=4/3 \pi abc$, where a, b, and c are the triaxial radii.}
\end{deluxetable*}

Below the synchronous orbit, progressive orbital decay inevitably brings the satellite within the Roche limit, where tidal stresses surpass its self-gravity and render the body gravitationally unbound, leading to tidal disruption. 
The classical fluid Roche radius is
\begin{equation}
R_\mathrm{Roche} = 2.456\left(\frac{\rho_\mathrm{p}}{\rho_\mathrm{s}} \right)^{1/3} R_\mathrm{p},
\label{eq:Roche}
\end{equation}
where $\rho_\mathrm{p}$ and $\rho_\mathrm{s}$ are the bulk densities of the primary and satellite, respectively, and $R_\mathrm{p}$ is the primary’s radius \citep{Murray1999}. 
For the Quaoar-Weywot system, based on the parameters listed in Table~\ref{tab:physparam}, $R_\mathrm{Roche}\simeq 1620~\mathrm{km}\simeq 2.97~R_\mathrm{p}\simeq1.5~D_\mathrm{p}$.
 
\begin{figure}
    \centering
    \includegraphics[width=1\linewidth]{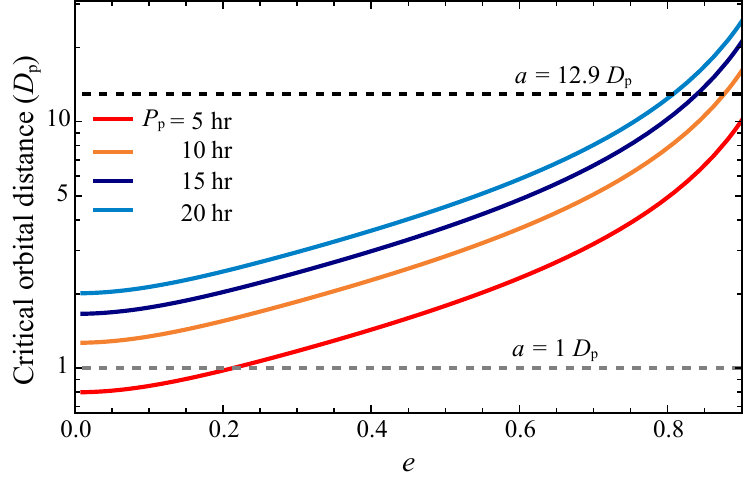}
    \caption{The critical orbital distance calculated analytically in the CTL framework as a function of Weywot’s orbital eccentricity for four assumed spin periods of Quaoar (5, 10, 15, and 20 hours). Beyond the critical distance, Weywot’s orbit expands, whereas below this threshold it decays.}
    \label{fig:critdist}
\end{figure}

The direction of the semimajor axis evolution of a prograde satellite orbit is determined by the ratio of the primary’s spin rate $\Omega_\mathrm{p}$ to the orbital mean motion $n$.
In particular, within the framework of the constant time-lag (CTL, e.g.,   \citealp{Mignard1979M&P....20..301M,Hut1981}), it can be shown analytically that
\begin{equation}
    \frac{\Omega_\mathrm{p}}{n} \gtrless \frac{N_a(e)}{N(e)},
\end{equation}
where
\begin{equation}
N_a(e) = \frac{1 + \tfrac{31}{2} e^{2} + \tfrac{255}{8} e^{4} + \tfrac{185}{16} e^{6} + \tfrac{25}{64} e^{8}}{(1-e^{2})^{15/2}},
\end{equation}
\begin{equation}
N(e) = \frac{1 + \tfrac{15}{2} e^{2} + \tfrac{45}{8} e^{4} + \tfrac{5}{16} e^{6}}{(1-e^{2})^{6}}
\end{equation}
(see, e.g., \citealp{NerondeSurgy1997A&A...318..975N,Laskar1997A&A...317L..75L,Levrard2007A&A...462L...5L,Correia2010Icar..205..338C,Leconte2010A&A...516A..64L}).
If $\Omega_\mathrm{p}/n \leq N_a(e)/N(e)$, tides raised on the primary extract angular momentum from the orbit and orbital decay occurs.
Conversely, if $\Omega_\mathrm{p}/n > N_a(e)/N(e)$, the tidal torque transfers angular momentum from the primary’s spin to the orbit and orbital expansion occurs.
Therefore, for a given orbital eccentricity, there is a critical orbital distance at which $\Omega_\mathrm{p}/n=N_a(e)/N(e)$; below this distance the orbit decays, whereas beyond it the orbit expands.
Examples of the critical orbital distance as a function of eccentricity calculated in the CTL framework for the Quaoar-Weywot system are presented in Figure~\ref{fig:critdist}, where four different spin periods are assumed for Quaoar.
Nevertheless, tidal interactions simultaneously act to damp the eccentricity. 
As a consequence, the critical radius is shifted inward, so that once the orbit is nearly circular, the satellite inevitably evolves outward.

Note that within the CPL and ACT frameworks, the direction of semimajor axis evolution is obtained from the sign of Equations~(\ref{eq:dadt}) and (\ref{eq:BEdadt}), respectively. 
For a given ($e,~\Omega_\mathrm{p}$), the critical orbital distance $a_\mathrm{crit}(e,~\Omega_{p}$) is the numerical root of $\dot{a}(a,~e,~\Omega_\mathrm{p})=0$, where the high-order $E^2_{2,0,k}(e)$ and $E^2_{2,1,k}(e)$ are computed by the method presented in  Appendix~\ref{apx:E2k}.
Therefore, modeling the complex orbital evolution of satellites in CPL or ACT framework necessitates the simultaneous numerical integration of orbital distance, eccentricity, and the spin rates of the interacting bodies.

A satellite on a retrograde orbit induces tidal interactions that act to decelerate both the planet's rotation and the satellite’s orbital motion, independent of the orbital eccentricity \citep{GoldreichSoter1966Icar....5..375G}. 
As a result, the satellite’s orbit decays irrespective of the location of the synchronous orbit. 
In the CPL framework, the characteristic timescale for orbital decay in the retrograde case is given by:
\begin{equation}
\tau_\mathrm{ret} = \left[ q \frac{3 }{2} n \frac{k_{2}}{Q} \left( \frac{R_\mathrm{p}}{a} \right)^5 \right]^{-1},
\end{equation}
where $q=M_\mathrm{s}/M_\mathrm{p}$ is the mass ratio, $n$ is the satellite’s mean motion, $k_2$ is Quaoar's Love number, $Q$ is the tidal quality factor.
Using the parameters listed in Table~\ref{tab:physpar} and adopting a plausible tidal dissipation factors, $10^{-4}\leq k_2/Q\leq 10^{-3}$, we estimate that retrograde satellites initially orbiting within $\sim20~D_\mathrm{p}$ would have decayed and been lost over the age of the Solar System.

Note that impact-generated satellites around asteroids are expected to form predominantly in prograde, near-equatorial orbits, because the reaccreting debris inherits the collision’s net angular momentum and aligns with the spin of the largest remnant \citep{Durda2004Icar..170..243D}. 
Retrograde birth (with respect to the final primary spin) is nevertheless possible when the target’s pre-impact rotation is comparable to and anti-aligned with the impact-delivered angular momentum, a regime in which impacts into rotating targets can drain spin angular momentum and despin or reorient the remnant before disk formation \citep{Sevecek2019A&A...629A.122S}.
An initially prograde configuration can also become apparently retrograde if the primary’s spin axis is reoriented after formation by the YORP effect over Myr-Gyr timescales \citep{Bottke2006AREPS..34..157B,Durech2018A&A...609A..86D}. 

In this work, we adopt the assumption that prograde satellites of asteroids originate from giant impacts, as proposed by \citet{Arakawa2019NatAs...3..802A}. Such collisions are expected to generate satellites at orbital distances of several primary diameters, with post-collisional orbits frequently exhibiting substantial eccentricities as a consequence of the dynamical excitation inherent to the formation process.
Nevertheless, a detailed investigation of retrograde satellites is deferred to future work.

When a satellite becomes tidally locked, its spin period becomes equal to its orbital period.
In the case of mutual tidal locking, the spin period of the primary also synchronizes with the satellite’s orbital period, so that both bodies continually present the same hemisphere to each other.
In this equilibrium state, tidal torques are effectively balanced, and the system enters a stable dynamical state in which further evolution is minimal unless acted upon by external perturbations \citep{Murray1981, Hut1981}.
Provided the system remains dynamically isolated, it may asymptotically approach this equilibrium configuration \citep{Levison2008}.
We note that, although tidal locking typically occurs at an early stage of tidal evolution, mutual locking did not occur in any of our simulations.

We adopt the classical fluid Roche limit (see Equation~\ref{eq:Roche}), thus the numerical integration is terminated when the satellite’s pericenter distance falls below this threshold, i.e., when $a(1 - e) < R_\mathrm{Roche}$.
The simulation is also terminated if the primary reaches its rotational break-up limit, defined by:
\begin{equation}
P_\mathrm{min} = 2\pi \left( \frac{G M_\mathrm{p}}{R_\mathrm{p}^3} \right)^{-1/2},
\end{equation}
which corresponds to a critical spin period of $\sim2.48$~hr for Quaoar, based on its adopted physical parameters.

The flattening (or oblateness) of a rotating body is defined as $f = (R_\mathrm{eq} - R_\mathrm{po}) / R_\mathrm{eq}$, 
where $R_\mathrm{eq}$ and $R_\mathrm{po}$ are the equatorial and polar radii, respectively. 
Quaoar has an equatorial radius of $R_\mathrm{eq} \simeq 643$~km 
and a polar radius of $R_\mathrm{po} \simeq 540$~km, resulting in a flattening of $f = 0.16$ \citep{Kiss2024AA...684A..50K}.
Assuming hydrostatic equilibrium and a homogeneous interior without cohesional force, the relatively small flattening is approximately related to the planet’s 
spin period by
\begin{equation}
    f \simeq C_{f} \left( \frac{2\pi}{P_\mathrm{p}} \right)^2 \frac{R_\mathrm{eq}^3}{G M_\mathrm{p}},
\end{equation}
see, e.g., \citet{Murray1999}.
If the feedback effect of rotational distortion on the planet’s gravity field -- arising from its departure from spherical symmetry -- is neglected, one obtains $C_f=1/2$, corresponding to a spin period of $P_\mathrm{p}\simeq5.62$~hr.
When the gravitational potential of the oblate body is taken into account, $C_f=5/4$, yielding $P_\mathrm{p}\simeq8.89$~hr.
In this study, we implicitly assume that Weywot originated from a giant impact, which melted and spun up Quaoar to the rotation period inferred from its present-day oblateness.
It should be emphasized, however, that Quaoar’s observed spin period is substantially longer, likely due to tidal evolution or other, as yet unidentified, processes; these possibilities are examined in detail in Section~\ref{sec:results}.

Since Quaoar exhibits a measurable departure from spherical symmetry, with a non-negligible flattening, it is necessary to assess the significance of this effect.
In this work, we treat CPL and ACT framework strictly as dissipative prescriptions; neither, by itself, encodes the primary’s permanent shape. 
The primary’s oblateness is a conservative quadrupole (characterized primarily by $J_2$) that modifies apsidal and nodal precession, $\dot \omega$ and $\dot \Omega$, without directly dissipating energy \citep{Kaula1966tsga.book.....K,Murray1999}. 
In the Darwin-Kaula expansion, these precessions enter the tidal forcing spectrum through the mode frequencies $\omega_{lmpq}$
so a nonzero $J_2$ shifts the entire tidal spectrum that CPL or ACT framework acts upon \citep{Kaula1964RvGSP...2..661K,Kaula1966tsga.book.....K,Efroimsky2013ApJ...764...26E}. 
Conceptually, this matters for spin equilibria and semimajor axis evolution: in CTL framework the near-synchronous fixed point at small $e$ is displaced from $\Omega\simeq n$ toward $\Omega \simeq n+\dot \omega$ \citep{Hut1981}, whereas in CPL the torque is piecewise in $\Omega$ and stationary states are not guaranteed -- oblateness changes which Fourier modes dominate and where sign changes occur \citep{Ferraz-Mello2013CeMDA.116..109F,Efroimsky2013ApJ...764...26E}. 
For $f=0.16$, the primary’s shape plays a significant role in orbital precession but has only a minor influence on tidal torques at Weywot’s distance. Thus, $J_2$ can generally be neglected in the tidal evolution terms ($\dot a,~\dot e,~\dot\Omega_\mathrm{p|s}$) when the objective is to constrain $k_2/Q$. It becomes important only when precise apsidal or nodal precession rates are required, when locating spin–orbit equilibria, or when evaluating possible resonances.

The spin period of Weywot remains undetermined, both at the time of its formation and in the present day. 
In the following analysis, we assume that Weywot is rotating in a prograde direction, with an initial spin period of approximately $P_\mathrm{s,ini}=15$~hr.
It is noteworthy that, regardless of Weywot’s initial spin period, tidal dissipation leads to rapid spin–orbit synchronization of the satellite within $\sim1$~Myr in both CPL and ACT framework. At the time of synchronization, Weywot’s spin period becomes equal to its orbital period, $\sim100$~hr, which subsequently lengthens as the orbit expands.

In the CPL framework, the evolution of the satellite's orbital distance is governed by the satellite-to-primary mass ratio, $q$, the Love numbers of the primary and satellite ($k_{2,\mathrm{p}}$ and $k_{2,\mathrm{s}}$), and their respective tidal dissipation functions ($Q_\mathrm{p}$ and $Q_\mathrm{s}$). 
For simplicity, the tidal dissipation function is assumed to be constant and identical for both bodies, such that $ k_\mathrm{2,p}/Q_\mathrm{p} = k_{2,s}/Q_\mathrm{s} = k_2/Q $. 
Note that $k_\mathrm{2,s}/Q_\mathrm{s}$ has negligible influence on the evolution of the orbital elements, owing to Weywot’s small size and mass.
The value of $k_2/Q$ is determined by fitting the observed semimajor axis of Weywot, as discussed in Section~\ref{sec:orbdistfit}.

In the ACT framework, we adopt the rigidity values following \citet{Fraser2013Icar..222..357F}: 
for Quaoar, the rigidity is taken as $\mu_\mathrm{p} = 4 \times 10^{10}~\mathrm{Pa}$, while for Weywot, we assume a rubble-pile structure with a rigidity of $\mu_\mathrm{s} = 4 \times 10^{9}~\mathrm{Pa}$.
Despite possible differences in the internal structures of Quaoar and Weywot, we adopt the simplifying assumption of equal viscosities, $\eta_\mathrm{p}=\eta_\mathrm{s}=\eta$.
Note that $\eta_\mathrm{s}$ exerts only a negligible influence on the overall tidal evolution of the system.

\section{Results}
\label{sec:results}

Section 3 presents the outcomes of our 4.5~Gyr orbit-averaged integrations using both the CPL and ACT frameworks, focusing on how tidal dissipation in Quaoar shapes Weywot’s semimajor axis, eccentricity damping, and the primary’s spin evolution. We first constrain the effective tidal parameters -- $k_2/Q$ in CPL and the mantle viscosity $\eta$ in ACT -- by requiring consistency with Weywot’s present orbital distance, then map the dependence of the secular evolution on the initial conditions ($a_{\rm ini}/D_p,\,e_{\rm ini},\,P_{p,{\rm ini}}$) across a broad parameter space.

\subsection{Constraining Tidal Dissipation Parameters}
\label{sec:orbdistfit}

\begin{figure*}
    \centering
    \includegraphics[width=0.99\textwidth]{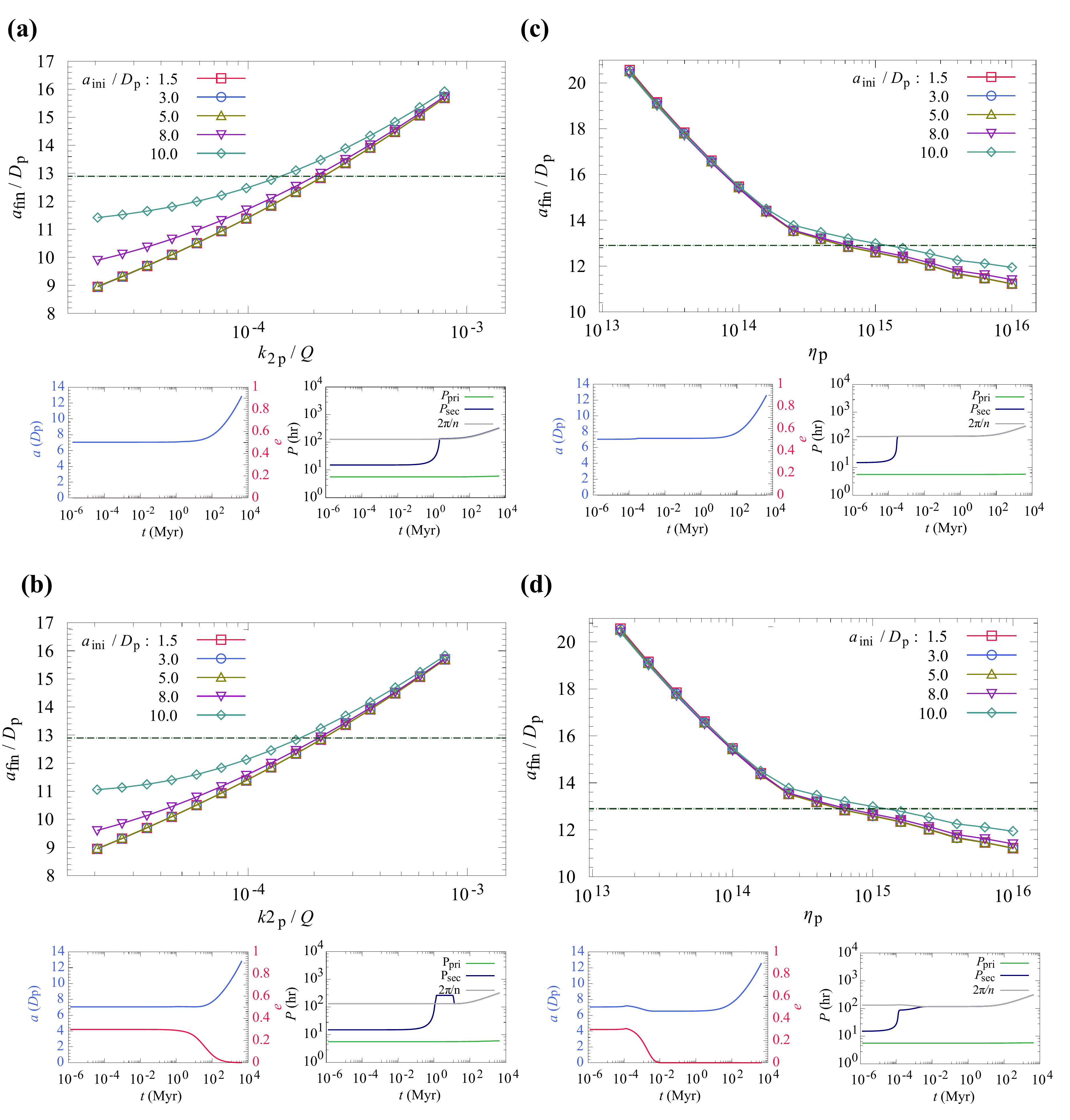}
    \caption{Final orbital distance of the satellite, $a_\mathrm{fin}/D_\mathrm{p}$, after 4.5 Gyr of tidal evolution for various initial orbital distances with $a_\mathrm{ini}/D_\mathrm{p} < 12.9$.
    Panels (a) and (b): $a_\mathrm{fin}/D_\mathrm{p}$ as a function of $k_{2}/Q$ using the CPL framework for initial eccentricities $e_\mathrm{ini} = 0$ and $e_\mathrm{ini} = 0.3$, respectively.
    Panels (c) and (d): $a_\mathrm{fin}/D_\mathrm{p}$ as a function of viscosity $\eta_\mathrm{p}$ using the ACT framework for the same eccentricities.
    The dotted green line represents Weywot's current orbital distance.
    Subpanels display the time evolution of the semimajor axis and eccentricity (bottom left subpanels) and the spin rates (bottom right subpanels), for models initialized with $a_\mathrm{ini}/D_\mathrm{p} = 7.0$.}
    \label{fig:findQ}
\end{figure*}

To construct models consistent with the observed orbital distance of Weywot, we performed tidal evolution simulations under the assumptions that its initial eccentricity is $e_\mathrm{ini} = 0$ and semimajor axis satisfies $a_\mathrm{ini}/D_\mathrm{p} < 12.9$.

Panel (a) of Figure~\ref{fig:findQ} shows the final orbital distance calculated in the CPL framework as a function of $k_{2}/Q$, along with the temporal evolution of the semimajor axis, eccentricity, and the spin period of Quaoar. 
As illustrated, the observed orbital distance of Weywot, $a_\mathrm{fin}/D_\mathrm{p} = 12.9$, can be reproduced by adopting a tidal parameter ratio of $k_{2}/Q \simeq 1.5-2 \times 10^{-4}$. 

In a subsequent set of models, the initial eccentricity of the satellite is set to $e_\mathrm{ini} = 0.3$, as illustrated in panel (b) of Figure~\ref{fig:findQ}. 
In this case, Weywot reaches synchronous rotation at a later time, approximately $t_\mathrm{sync} \simeq 2$~Myr. 
Notably, the eccentricity is efficiently damped, reaching near-zero values within roughly $10^{-2}$~Myr.
As a result, the previously determined $k_{2}/Q \simeq 1.5-2 \times 10^{-4}$ remains consistent with the observed orbital configuration, further validating the robustness of the model across different initial eccentricities.

The previous experiments were repeated using the ACT framework. 
In this framework, tidal evolution is governed by the viscosity of the body through the creep function, rather than by its rigidity (see Equation~\ref{eq:creep}). Accordingly, in panels (c) and (d) of Figure~\ref{fig:findQ}, we plot the final orbital distance normalized to the primary diameter, $a_\mathrm{fin}/D_\mathrm{p}$, as a function of the primary's viscosity, $\eta_\mathrm{p}$.
As shown, the results are broadly consistent with those obtained from the CPL framework; in particular, the final orbital distance exhibits minimal sensitivity to the initial semimajor axis, $a_\mathrm{ini}/D_\mathrm{p}$.
However, tidal processes proceed significantly more rapidly in the ACT framework. 
Specifically, eccentricity damping and rotational synchronization occur on much shorter timescales, with $t_\mathrm{damp} \simeq 10^{-2}$~Myr and $t_\mathrm{sync} \simeq 10^{-3}$~Myr, respectively.
The viscosity value that best reproduces the observed orbital distance of Weywot in the ACT framework is found to be $\eta \simeq 6\times10^{14}~\mathrm{Pa\,s}$.

Since the satellite’s rotation is rapidly synchronized with its orbital period and eccentricity is damped in both the CPL and ACT frameworks, its role in the subsequent long-term tidal evolution is negligible. 
Consequently, the secular tidal evolution reduces to the tide raised on (and dissipated within) the primary.
It is also important to note that, in both the CPL and ACT frameworks, Quaoar’s spin period remains effectively unchanged from its initial value throughout the simulations.
This outcome is in clear contradiction with the observed spin period of $P_\mathrm{p}=17.7$~hr, highlighting a fundamental limitation of the current tidal models or the history of the formation of Weywot.
This discrepancy will be addressed and discussed in detail in the following.

\subsection{Testing Tidal Models Against Initial Conditions}
\label {sec:lps}

In the previous section, we demonstrated that the observed orbital distance of Weywot can be reproduced assuming $1.5 \leq a_\mathrm{ini}/D_\mathrm{p}\leq10$ and zero or moderate initial eccentricity of $e = 0.3$, assuming $k_{2}/Q = 1.5 \times 10^{-4}$ (CPL) or $\eta = 6\times10^{14}~\mathrm{Pa\,s}$ (ACT). 
In this section, we extend our parameter study to explore a wider range of initial orbital distances and eccentricities.
To assess the influence of the initial semimajor axis on the tidal evolution, we sample 20 values of $a_\mathrm{ini}/D_\mathrm{p}$ logarithmically distributed over the range $1.5 \leq a_\mathrm{ini}/D_\mathrm{p} \leq 100$. 
We further investigate the effect of the Wewot’s initial orbital eccentricity by sampling 10 uniformly spaced values over the range $0 \leq e_\mathrm{ini} \leq 0.9$.

\begin{figure*}
    \centering
    \includegraphics[width=0.49\linewidth]{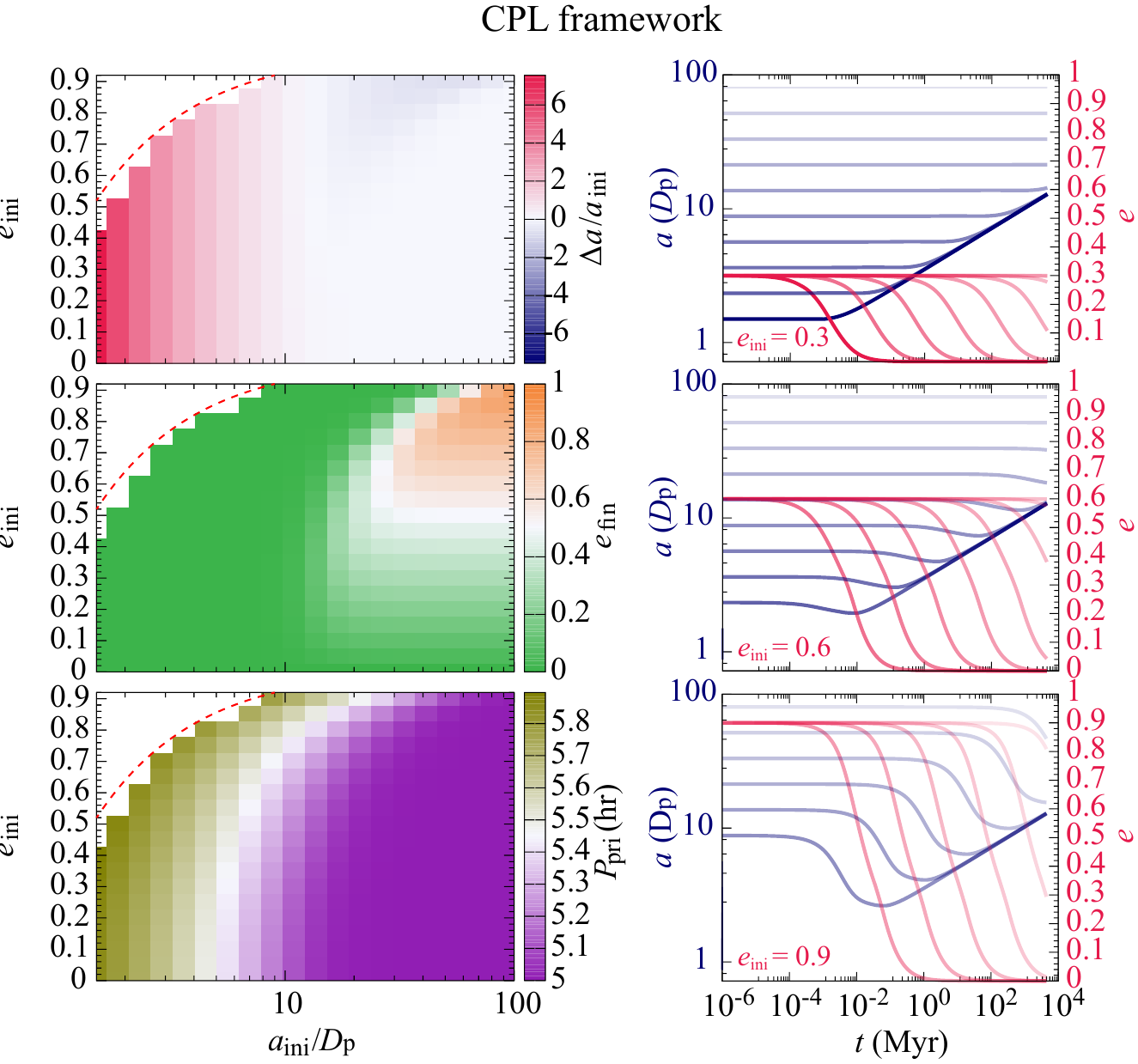}
    \includegraphics[width=0.49\linewidth]{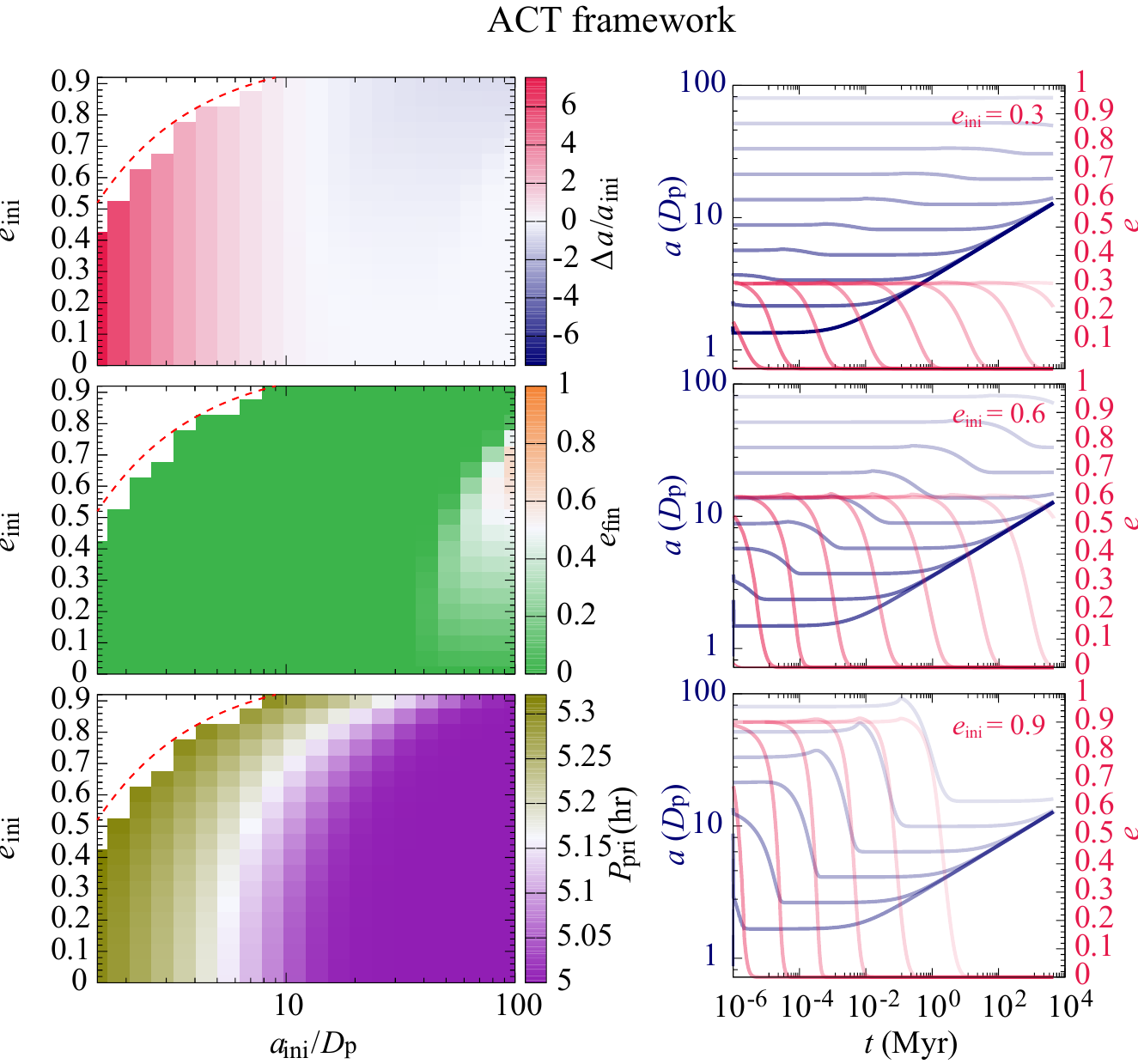}
    \caption{Results of 4.5 Gyr tidal evolution simulations for the Quaoar-Weywot system, assuming an initial spin period of the primary $P_\mathrm{p} = 5$ hr. The CPL and ACT frameworks are shown in the left and right columns, respectively.
    The top row of density plots displays the relative change in the satellite's orbital distance; blue indicates orbital decay, red indicates expansion.
    The middle row shows the final orbital eccentricity of the satellite.
    The bottom row presents the final spin period of the primary.
    The red dashed curve marks the Roche limit for each $(a_\mathrm{ini}, e_\mathrm{ini})$ pair, while white regions correspond to cases where the satellite crosses this limit.
    Examples of the time evolution of the semimajor axis (blue curves) and eccentricity (red curves) are shown for selected initial eccentricities ($e_\mathrm{ini} = 0.3$, $0.6$, and $0.9$) and orbital distances, with lighter colors indicating higher $a_\mathrm{ini}$ values.}
    \label{fig:map-5hr}
\end{figure*}

Figure~\ref{fig:map-5hr} shows the results for an initial Quaoar spin period of $P_\mathrm{p,ini}=5$~hr, computed using both the CPL and ACT frameworks.
As shown, both frameworks yield qualitatively similar outcomes; however, eccentricity damping occurs significantly more rapidly in the viscoelastic ACT framework.

The outset that all models predict either orbital expansion or orbital decay, depending on the initial semimajor axis and eccentricity.
Specifically, models with low initial eccentricity ($e_\mathrm{ini}\lesssim0.3$) and $a_\mathrm{ini}/D_\mathrm{p}\lesssim 10$ exhibit only orbital expansion, whereas models with higher initial eccentricity ($e_\mathrm{ini}\gtrsim 0.3$) and $a_\mathrm{ini}/D_\mathrm{p}\gg 10$ undergo an initial phase of orbital decay followed by subsequent expansion.
Orbital expansion arises because eccentricity is damped while the critical orbital distance decreases, allowing tidal interactions to drive the orbit outward (see Figure~\ref{fig:critdist}).
In practice, however, over 4.5~Gyr of tidal evolution these cases exhibit little to no net orbital expansion, with $\Delta a/a_\mathrm{ini}\simeq 0$.

Two qualitatively distinct dynamical regimes emerge from the simulations. 
In group (a), the satellite's eccentricity is efficiently damped within 4.5~Gyr (green regions in the middle panels of Figure~\ref{fig:map-5hr}), while in group (b), tidal evolution over the same timescale is insufficient to significantly reduce the eccentricity (orange regions in the middle panels of Figure~\ref{fig:map-5hr}). 
Group (a) is generally associated with systems having $a_\mathrm{ini}/D_\mathrm{p} \lesssim 20$ in the CPL framework, while in the ACT framework it extends to much wider systems with $a_\mathrm{ini}/D_\mathrm{p} \lesssim 60$. 
This difference arises from the significantly stronger eccentricity damping in the ACT framework.
Although the highly eccentric cases in group (b) are capable of reproducing Weywot’s current orbital distance, the observed low eccentricity is better matched by models calculated in the ACT framework.
It is also noteworthy that none of the models reproduce Quaoar’s observed spin period (bottom panels of Figure~\ref{fig:map-5hr}), highlighting a limitation inherent to both tidal prescriptions or to the assumed formation scenario of Weywot.

\begin{figure*}
    \centering
    \includegraphics[width=1\linewidth]{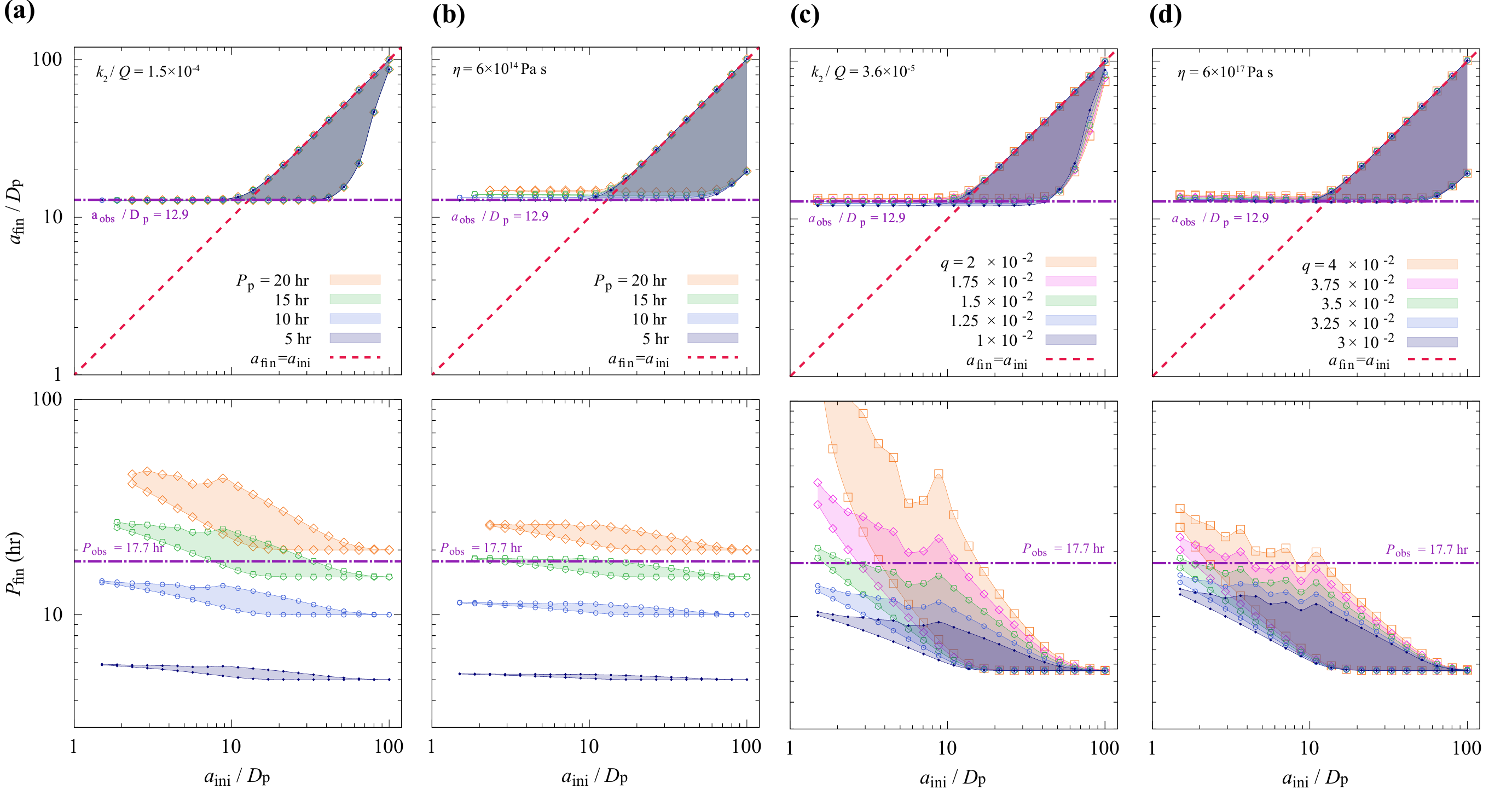}
    \caption{Final orbital distance of Weywot ($a_\mathrm{fin}/D_\mathrm{p}$, top panels) and final spin period of Quaoar ($P_\mathrm{p,fin}$, bottom panels) after 4.5 Gyr of tidal evolution, shown as a function of the initial orbital distance ($a_\mathrm{ini}/D_\mathrm{p}$).
    The tidal evolution shown in panels (a) and (c) is calculated within the CPL framework, whereas panels (b) and (d) are computed using the ACT framework.
    Panels (a) and (b) show results computed for four different initial rotation periods of Quaoar: $P_\mathrm{p,ini} = 5$, 10, 15, and 20 hr (panels a and b).
    Panels (c) and (d) show results for four different mass ratio of Quaoar and Weywot.
    The dashed black line marks the locus of unchanged orbital distance ($a_\mathrm{fin} = a_\mathrm{ini}$), indicating cases with no net orbital evolution.
    The observed values of Weywot’s current orbital distance ($a = 12.9~D_\mathrm{p}$) and Quaoar’s spin period ($P_\mathrm{p} = 17.7$ hr) are shown as magenta dot-dashed lines for reference.}
    \label{fig:a-a-comp}
\end{figure*}

Three additional models are examined in which Quaoar's initial spin period is set to $P_\mathrm{p,ini} = 10$, $15$, and $20$~hr. 
Panels (a) and (b) of Figure~\ref{fig:a-a-comp} display the final-to-initial orbital distance ratio ($a_\mathrm{fin}/D_\mathrm{p}$ vs. $a_\mathrm{ini}/D_\mathrm{p}$) and the final spin period of Quaoar ($P_\mathrm{p,fin}$ vs. $a_\mathrm{ini}/D_\mathrm{p}$) for the CPL and ACT frameworks, respectively. 
The plotted symbols indicate the range (maximum and minimum) of outcomes obtained for a given set of initial conditions: $a_\mathrm{ini}/D_\mathrm{p}$, $e_\mathrm{ini}$, and $P_\mathrm{p,ini}$.
The maxima and minima of $a_\mathrm{fin}/D_\mathrm{p}$ generally correspond to small and large initial eccentricities, respectively.
Overall, the results suggest that a suitable choice of tidal parameters -- $k_{2}/Q$ for the CPL framework or $\eta_\mathrm{p}$ for the ACT framework -- can reproduce the observed orbital distance of Weywot across a wide range of initial semimajor axes and eccentricities independent of $P_\mathrm{p}$.
Notably, initially high eccentric and very wide $10\ll a_\mathrm{ini}/D_\mathrm{p}\lesssim50$ models exhibiting orbital decay  are also capable of matching the observed orbit of Weywot, however, it requires a relatively large initial eccentricity.

The observed spin period of Quaoar, $17.7$~hr, can be reproduced only if the initial spin period is assumed to be $P_\mathrm{p,ini} \simeq 15$~hr. 
This result reflects the relatively weak tidal torque exerted by Weywot -- given the low satellite-to-primary mass ratio ($q = 3.5 \times 10^{-3}$) -- which is insufficient to produce a significant change in Quaoar’s spin rate over the system’s evolutionary timescale.
However, we emphasize that such a slow rotation is inconsistent with Quaoar’s observed oblateness, suggesting a discrepancy between its current shape and rotational state that cannot be explained by tidal evolution alone.

To address the discrepancy between Quaoar’s observed spin period and that predicted by tidal evolution with Weywot's nominal mass, we conducted an extended parameter study assuming larger satellite masses. 
In this scenario, the tidal torque can become sufficiently strong to spin down Quaoar from an initial spin period to the observed value of  $17.7$~hr.
The results assuming $P_\mathrm{ini}=5$~hr are presented in panels (c) and (d) of Figure~\ref{fig:a-a-comp}, corresponding to the CPL and ACT frameworks, respectively. 
Following a preliminary assessment of the possible mass ratio between Quaoar and Weywot, we investigated satellite-to-primary mass ratios in the range $1 \times 10^{-2} \leq q \leq 2 \times 10^{-2}$ within the CPL framework, while slightly higher values, $3 \times 10^{-2} \leq q \leq 4 \times 10^{-2}$, were adopted for the ACT framework.
The tidal dissipation parameters were set to $k_{2}/Q = 3.6 \times 10^{-5}$ for the CPL framework and $\eta = 6 \times 10^{16}$~Pa~s for the ACT framework to ensure consistency with the observed orbital distance of Weywot.
The results indicate that a range of initial orbital configurations could reproduce both the observed semimajor axis of Weywot and Quaoar’s slow rotation. 
Our additional simulations show that for $P_\mathrm{ini}=8.9$~hr, slightly smaller -- yet still more than half an order of magnitude larger -- mass ratios are required to account for Quaoar’s spin-down.
However, achieving this outcome requires fine-tuning of Weywot’s initial eccentricity, since Quaoar’s final spin period matches the observed value only for a specific values of initial eccentricity and semimajor axis.
Note that the satellite masses required in these models exceed observational estimates of Weywot’s mass by approximately an order of magnitude, raising questions about the physical plausibility of such scenarios.

\section{Summary and Conclusions} 
\label{sec:conclusions}

In this paper, we presented numerical integrations to investigate the tidal evolution of the Quaoar-Weywot system over a timescale of $4.5$~Gyr. Two tidal frameworks were employed: a constant phase-lag (CPL) and an Andrade creep-tide (ACT) framework, both under the assumption of orbit-averaged evolution; see details in Appendices~\ref{sec:const-lag} and  \ref{sec:nonconst-lag}.
To accurately handle high orbital eccentricities, both frameworks incorporate numerical expansions of the eccentricity functions $E_{2,0,k}(e)$ and $E_{2,1,k}(e)$, pre-computed over range $-200 \leq k \leq 200$. 
These functions were evaluated at 10,000 discrete eccentricity values to ensure both precision and computational efficiency; see details in Appendix~\ref{apx:E2k}.

Assuming a Weywot mass consistent with recent observational constraints (see, e.g., \citealp{Kiss2024AA...684A..50K} and references therein), the tidal torque exerted by the satellite is insufficient to achieve mutual synchronization between Quaoar’s spin and Weywot’s orbital period.
This implies that, although tidal evolution is still ongoing, its present-day effect is negligible.
Consequently, the current orbital configuration of the Quaoar–Weywot system can be considered quasi-stationary.
After 4.5 Gyr of tidal evolution, both the CPL and ACT frameworks yield a Weywot orbital distance consistent with current observations. This requires either $a_\mathrm{ini}/D_\mathrm{p}\lesssim10$ with arbitrary initial eccentricity, or $10\lesssim a_\mathrm{ini}/D_\mathrm{p}\lesssim 50$ with relatively high initial eccentricity $e\gtrsim0.6$. 
In all models that satisfy these conditions, the orbital eccentricity is damped to effectively zero, within the numerical accuracy of the integrations.
In the CPL framework, adopting a tidal parameter of $k_{2}/Q \simeq 1.5 \times 10^{-4}$ allows reproduction of Weywot’s observed orbital distance of $12.9~D_\mathrm{p}$. 
Similarly, in the ACT framework, this orbital configuration can be matched by assuming a viscosity of $\eta \simeq 6 \times 10^{14}$~Pa~s.

Quaoar’s bulk density of $1.67–1.77~\mathrm{g~cm^{-3}}$ is strikingly intermediate between that of pure water ice ($\simeq0.93~\mathrm{g~cm^{-3}}$ at Kuiper‑belt temperatures) and anhydrous silicate rock ($\simeq3.0~\mathrm{g~cm^{-3}}$). 
Applying a two‑component mass balance with a small allowance for residual porosity ($<5~\%$) yields a rock mass fraction of roughly $50–60\%$. 
Densities in this range are common among the largest KB objects and are generally interpreted as evidence for partial differentiation into a rocky core overlain by an $\mathrm{H_2 O}$‑rich mantle \citep{Barr2016MNRAS.460.1542B}.
Thermal–orbital models show that short‑lived radionuclides and long‑term accretional heating can readily melt internal ice in $\sim1000$~km bodies, driving segregation of silicates toward the center \citep{McKinnon2017Icar..287....2M}.
This is also supported by JWST NIRSpec measurements \citep{Emery2024Icar..41416017E}.
Consistent with that picture, the tidal parameters inferred for Quaoar point to a visco‑elastic interior that is neither purely icy nor monolithically rocky. 
The tidal dissipation parameter $k_2/Q\simeq1.5\times10^{-4}$ (CPL framework) lies near values measured or modeled for large icy satellites such as Rhea or Dione \citep{Lainey2020NatAs...4.1053L}, indicating appreciable but not extreme dissipation. 
Likewise, a creep‑tide viscosity of $\eta\simeq6\times10^{14}~\mathrm{Pa~s}$ (ACT framework) matches laboratory estimates for water ice in the 150–180~K range \citep{Goldsby2001LPI....32.2067G} and is several orders of magnitude lower than that expected for silicates, implying that the torque is predominantly absorbed in a warm, ductile ice layer rather than in the core.

Thermochemical modelling further indicates that small amounts of $\mathrm{N H_3}$ or $\mathrm{CH_4}$ clathrate, if present, would marginally lower the mantle viscosity but would have little impact on the bulk density (see, e.g., \citealp{Hussmann2006Icar..185..258H,Courville2023PSJ.....4..179C}). 
Conversely, a rock‑dominated body ($\rho>2~\mathrm{g~cm^{-3}}$) can be ruled out, and an ice‑dominated “rubble pile” would possess a much larger $k_2/Q$ and far lower $\eta$ than suggested.
Thus, the most plausible composition for Quaoar is a partially differentiated dwarf planet containing roughly equal masses of silicate rock and $\mathrm{H_2O}$‑dominated ices, with an ice shell warm enough at depth to creep on gigayear timescales yet sufficiently stiff near the surface to sustain the observed shape and oblateness. 
Future detections of libration of Weywot's orbit -- e.g. via continued stellar‑occultation campaigns -- could further test this internal configuration by independently constraining the thickness and mechanical properties of Quaoar’s ice mantle \citep{Proudfoot2024AJ....167..144P}

According to \citet{Kiss2024AA...684A..50K}, both the thermal and optical light curves of Quaoar most likely support a double-peaked rotational solution, with a confidence level of approximately 98\%. Consequently, the alternative single-peaked solution, implying a faster rotation, is statistically disfavoured with only about a 2\% probability. Interestingly, when the gravitational potential of an oblate body is taken into account, the corresponding rotational flattening would be consistent with a spin period of 8.89 hours. Given that the satellite Weywot is dynamically incapable of significantly altering Quaoar’s rotation rate, this result remains compatible with the single-peaked interpretation.
Nevertheless, the presently favored spin period of Quaoar, $P = 17.7$~hr, is substantially longer than expected from tidal evolution models starting from an initial period of $5.6$~hr or $8.89$~hr, values that are compatible with its measured oblateness.
Reproducing such a rotational slowdown requires a substantially stronger tidal torque from Weywot, which in turn implies a satellite mass at least an order of magnitude greater than current observational estimates.
However, under this assumption, achieving the observed orbital distance of Weywot also necessitates a finely tuned initial orbital eccentricity for a given initial semimajor axis.
Thus, irrespective of rheology, Weywot’s tides are intrinsically too weak to despin Quaoar to $17.7$~hr over 4.5~Gyr. 

It is worth noting that the density estimates for the Quaoar--Weywot system have evolved considerably since its discovery. Reported values include 
\( 4.2~\mathrm{g~cm^{-3}} \) \citep{fraserbrown10}, 
\( 2.13~\mathrm{g~cm^{-3}} \) \citep{fornasieretal13,brownbutler17}, 
and \( 1.99~\mathrm{g~cm^{-3}} \) \citep{Pereira2023A&A...673L...4P}. 
All of these are significantly higher than the value adopted in the present study. 
Assuming a higher bulk density -- and hence a proportionally larger system mass 
(by approximately 20\% or 10\% for the latter two estimates, respectively )-- would lead to a somewhat faster tidal evolution. 
Consequently, the corresponding plausible values of \( k_2/Q \) or \( \eta \) required to reproduce Weywot’s current orbital separation would be lower by roughly 20\% or 10\%, respectively.


A model-independent diagnostic for the Quaoar-Weywot system
\begin{equation}
    \frac{\tau_{\Omega}}{\tau_a}\simeq0.8\,\frac{M_\mathrm{p}}{M_\mathrm{s}}\left(\frac{R_\mathrm{p}}{a}\right)^2\frac{\Omega_\mathrm{p}}{n}\simeq6,
\end{equation}
where $\tau_{\Omega}$ and $\tau_a$ is the characteristic time-scale of spin and orbital growth, respectively, shows that primary despinning proceeds several times more slowly than orbital expansion for the measured masses and semimajor axis.
Moreover, the spin angular-momentum change required to lengthen the spin of Quaoar exceeds Weywot’s entire present orbital angular momentum by several times \citep{Taylor2010CeMDA.108..315T,Fraser2013Icar..222..357F,Braga-Ribas2025RSPTA.38340200B}. 
Consistent pathways are therefore: (i) the $5.6$~hr or $8.89$~hr spin period of Quaoar inference from its oblateness is invalid because Quaoar’s present shape is non-hydrostatic/fossil, so the current $17.7$~hr spin is largely primordial; (ii) early coupling to a transient, more massive inner satellite or debris disk presumably on retrograde orbit that siphoned the requisite angular momentum and was later disrupted or removed; and/or (iii) a secondary giant impact  -- occurring retrograde with respect to Quaoar’s spin -- that directly established a comparatively slow post-impact rotation. 
As a final note, we emphasize that the multiple rings in the Quaoar–Weywot system \citep{Morgadoetal2023Natur.614..239M,Kiss2024AA...684A..50K,Nolthenius_2025} are not expected to play a significant role in the tidal evolution of the system, owing to their low mass and their transient nature under the influence of radiation pressure (see, e.g., \citealp{Regaly2025A&A...697A.116R} and references therein).
However, tidal evolution may influence Quaoar’s ring system, since the 6:1 mean-motion resonance between Quaoar and its satellite Weywot is thought to contribute to the ring’s confinement.
The satellite’s orbital eccentricity can generate an equilibrium region that concentrates ring material into an arc (see, e.g., \citealp{Pereira2023A&A...673L...4P,Morgadoetal2023Natur.614..239M}).
However, as tidal evolution progresses, the resonance location drifts and Weywot’s eccentricity decreases, potentially inhibiting long-term ring stabilization. This phenomenon merits further investigation in future studies.

\section*{Acknowledgments}
V. F. acknowledges financial support from the undergraduate research assistant program of the Konkoly Observatory and the EKÖP-24-2-I-ELTE-64 University Research Scholarship Program of the Ministry for Culture and Innovation
from the source of the National Research, Development and Innovation Fund.
The research leading to these results has received funding from the K-138962 and TKP2021-NKTA-64 grants of the National Research, Development and Innovation Office (NKFIH, Hungary).


%

\vspace{5mm}





\bibliography{Quaoar-Weywot}{}
\bibliographystyle{aasjournal}



\appendix




\section{Constant Phase-Lag Tide Model} 
\label{sec:const-lag}

In the constant phase-lag (CPL) framework, the tidal evolution of the Quaoar-Weywot system is treated in an orbit-averaged framework using the formulation of \citet{Ferraz-Mello2013CeMDA.116..109F}, combined with the classical CPL framework of \citet{GoldreichPeale1966AJ.....71..425G}. 
We implicitly assume that the orbital inclinations with respect to both the primary’s and the satellite’s equatorial planes are negligibly small. 
Under these assumptions, the rates of change of the satellite’s semimajor axis ($a$) and eccentricity ($e$) are given by:
\begin{eqnarray}
   \frac{da}{dt}&=&q\frac{3na}{2} \left(\frac{R_\mathrm{p}}{a}\right)^5 \frac{k_\mathrm{2,p}}{Q_\mathrm{p}}\sum_{k=-N}^{N}\left[(2-k)E^2_{2,0,k}(e)\mathrm{sgn}(2\Omega_\mathrm{p}-(2-k)n)\right]\nonumber\\
   &&+\frac{1}{q}\frac{3na}{2} \left(\frac{R_\mathrm{s}}{a}\right)^5 \frac{k_\mathrm{2,s}}{Q_\mathrm{s}}\sum_{k=-N}^{N} \left[(2-k)E^2_\mathrm{2,0,k}(e)\mathrm{sgn}(2\Omega_\mathrm{s}-(2-k)n)\right],
   \label{eq:dadt}
\end{eqnarray}
\begin{eqnarray}
   \frac{de}{dt}&=-q&\frac{3n}{4e} \left(\frac{R_\mathrm{p}}{a}\right)^5 \frac{k_\mathrm{2,p}}{Q_\mathrm{p}}\sum_{k=-N}^{N}\left[\left(2\sqrt{1-e^2}-(2-k)(1-e^2)\right)E^2_{2,0,k}(e)\mathrm{sgn}(2\Omega_\mathrm{p}-(2-k)n)\right]\nonumber\\
   &&-\frac{1}{q}\frac{3n}{4e} \left(\frac{R_\mathrm{s}}{a}\right)^5 \frac{k_\mathrm{2,s}}{Q_\mathrm{s}}\sum_{k=-N}^{N}\left[\left(2\sqrt{1-e^2}-(2-k)(1-e^2)\right)E^2_\mathrm{2,0,k}(e)\mathrm{sgn}(2\Omega_\mathrm{s}-(2-k)n)\right],
   \label{eq:dedt}
\end{eqnarray}
The rate of change in the spin of the primary ($\Omega_\mathrm{p}$) and the satellite ($\Omega_\mathrm{s}$) are as follows:
\begin{equation}
   \frac{d\Omega_\mathrm{p}}{dt}=-\frac{q^2}{1+q} \frac{15n^2}{4}\left(\frac{R_\mathrm{p}}{a}\right)^3 \frac{k_\mathrm{2,p}}{Q_\mathrm{p}}\sum_{k=-N}^{N}\left[E^2_{2,0,k}(e)\mathrm{sgn}\left(\Omega_\mathrm{p}-(2-k)n\right)\right],
   \label{eq:dOmegapdt}
\end{equation}
\begin{equation}
   \frac{d\Omega_\mathrm{s}}{dt}=-\frac{1}{q+q^2}\frac{15n^2}{4} \left(\frac{R_\mathrm{s}}{a}\right)^3 \frac{k_\mathrm{2,s}}{Q_\mathrm{s}}\sum_{k=-N}^{N}\left[E^2_{2,0,k}(e)\mathrm{sgn}\left(\Omega_\mathrm{s}-(2-k)n\right)\right],
   \label{eq:dOmegasdt}
\end{equation}
where $n$, $q$, $R_\mathrm{p}$ and $R_\mathrm{s}$ are the mean motion of the satellite, the satellite-to-primary mass ratio, and the primary and satellite radii, respectively.
In the above expressions, $q = M_\mathrm{s}/M_\mathrm{p}$ denotes the satellite-to-primary mass ratio. 
The tidal Love numbers of the primary ($k_{2,\mathrm{p}}$) and the satellite ($k_{2,\mathrm{s}}$) are given by:
\begin{equation}
    k_{2,\mathrm{p|s}} = \frac{3}{2} \left(1 + \mu_{\mathrm{eff,p|s}} \right)^{-1},
    \label{eq:k2}
\end{equation}
where the dimensionless effective rigidity \citep[e.g.,][]{Goldreich+2009,Cheng+2014} is defined as:
\begin{equation}
    \mu_{\mathrm{eff,p|s}} = \frac{38\pi}{3G} \frac{\mu_{\mathrm{p|s}} R_{\mathrm{p|s}}^4}{M_{\mathrm{p|s}}^2},
\end{equation}
with $\mu_\mathrm{p}$ and $\mu_\mathrm{s}$ being the rigidities of the primary and satellite, respectively, and $G$ the gravitational constant. 
Note that the expression of $k_2,p|s$ defined in Equation~(\ref{eq:k2}) corresponds to homogeneous incompressible body.
The parameters $Q_\mathrm{p}$ and $Q_\mathrm{s}$ represent the tidal dissipation functions of the primary and the satellite, respectively. 

The eccentricity function $E_{2,k}(e)$ arises from the Fourier expansion of the solution to Keplerian motion, as originally presented by \citet{Cayley1861MmRAS..29..191C}. 
A tabulated expansion of these functions for the range $-7 \leq k \leq 7$ is provided in the supplementary material of \citet{Efroimsky2012CeMDA.112..283E}. 
A similar approach was adopted by \citet{Arakawa2019NatAs...3..802A}, who truncated the summation in Equations~(\ref{eq:dadt})–(\ref{eq:dOmegapdt}) at $N = 3$.
However, as demonstrated by \citet{Ferraz-Mello2013CeMDA.116..109F}, the convergence of the Cayley series expansion is limited to moderate eccentricities ($e \lesssim 0.4$). 
To extend the applicability of our tidal evolution models to arbitrary eccentricities ($0 \leq e_\mathrm{ini} < 1$), we adopt a numerical approach to compute $E_{2,0,k}(e)$ with much broader range of $-200
\leq k \leq200$ values, following the methodology described in \citet{Ferraz-Mello2013CeMDA.116..109F}. 
Additional details are provided in Appendix~\ref{apx:E2k}.

\section{Andrade Creep-tide Model }
\label{sec:nonconst-lag}

Our Andrade creep-tide (ACT) framework is based on the formulation by \citet{Boue2019CeMDA.131...30B}, which incorporates the Andrade viscoelastic rheology. 
This approach was also adopted by \citet{Arakawa2021AJ....162..226A}, who coupled it with the thermal evolution of the system, accounting for both tidal heating and radiogenic decay.
In the present study, however, we neglect the thermal evolution and focus solely on the tidal dynamics. 
The rates of change in the semimajor axis and eccentricity of the Quaoar-Weywot system are given as follows:
\begin{equation}
   \frac{da}{dt}=q 2 n a \left(\frac{R_\mathrm{p}}{a}\right)^5 
   \sum_{k=-N}^{N} {\left[ \frac{3 {( 2 + k )}}{4} \mathcal{A}_{{\rm p}, k} + \frac{q}{4} \mathcal{B}_{{\rm p}, k} \right]}
   +\frac{1}{q}2na \left(\frac{R_\mathrm{s}}{a}\right)^5
   \sum_{k=-N}^{N} {\left[ \frac{3 {( 2 + k )}}{4} \mathcal{A}_{{\rm s}, k} + \frac{q}{4} \mathcal{B}_{{\rm s}, k} \right]},
   \label{eq:BEdadt}
\end{equation}
\begin{eqnarray}
   \frac{de}{dt}&=q&\frac{n}{e} \left(\frac{R_\mathrm{p}}{a}\right)^5 
   \sum_{k=-N}^{N} {\left[ \frac{3 {( 2 + k )} {( 1 - e^{2} )} - 6 \sqrt{1 - e^{2}}}{4} \mathcal{A}_{{\rm p}, k} + \frac{k}{4} {( 1 - e^{2} )} \mathcal{B}_{{\rm p}, k} \right]}\\
   &&+\frac{1}{q}\frac{n}{e} \left(\frac{R_\mathrm{s}}{a}\right)^5 
   \sum_{k=-N}^{N} {\left[ \frac{3 {( 2 + k )} {( 1 - e^{2} )} - 6 \sqrt{1 - e^{2}}}{4} \mathcal{A}_{{\rm s}, k} + \frac{k}{4} {( 1 - e^{2} )} \mathcal{B}_{{\rm s}, k} \right]},
   \label{eq:BEdedt}
\end{eqnarray}
The rate of change in the spin of the primary ($\Omega_\mathrm{p}$) and the satellite ($\Omega_\mathrm{s}$) are as follows:
\begin{equation}
   \frac{d\Omega_\mathrm{p}}{dt}=-\frac{q^2}{1+q} \frac{15n^2}{4}\left(\frac{R_\mathrm{p}}{a}\right)^3
   \sum_{k=-N}^{N} \mathcal{A}_{{\rm p}, k},
   \label{eq:BEdOmegapdt}
\end{equation}
\begin{equation}
   \frac{d\Omega_\mathrm{s}}{dt}=-\frac{1}{q+q^2}\frac{15n^2}{4} \left(\frac{R_\mathrm{s}}{a}\right)^3 
   \sum_{k=-N}^{N} \mathcal{A}_{{\rm s}, k},
   \label{eq:BEdOmegasdt}
\end{equation}
where
\begin{equation}
    \mathcal{A}_{{\rm p|s}, k}  =  {\left[ E_{2, 0, k} {( e )} \right]}^{2}\ {\rm Im}{\left[ k_{2, {\rm p|s}} {\left( {( 2 + k )} n - 2 \Omega_{\rm p|s} \right)} \right]},
    \label{eq:BEA}
\end{equation}
\begin{equation}
    \mathcal{B}_{{\rm p|s}, k}  =  {\left[ E_{2, 1, k} {( e )} \right]}^{2}\ {\rm Im}{\left[ k_{2, {\rm p|s}} {\left( k n \right)} \right]}.
    \label{eq:BEB}
\end{equation}
In Equations~(\ref{eq:BEA}) and (\ref{eq:BEB}), the eccentricity functions are
\begin{equation}
    E_{2,q,k}(e)=\frac{1}{2\pi\sqrt{1-e^2}}\int_0^{2\pi}\frac{1+e\cos(\nu)}{1-e^2}\cos((2-2q)\nu-(2-2q+k)l)d\nu,
    \label{eq:E2pk}
\end{equation}
where $q=0$ or $q=1$.
To compute the values of $E_{2,0,k}(e)$ and $E_{2,1,k}(e)$ numerically over the range $-200 \leq k \leq 200$, we employed Equations~(\ref{eq:mean-ano}) and (\ref{eq:beta}), utilizing the same series expansion approach as adopted in the CPL framework.

In Equations~(\ref{eq:BEA}) and (\ref{eq:BEB}), the complex Love number of the primary and the satellite are given as
\begin{eqnarray}
    \tilde{k}_\mathrm{2,p|s} {( \omega )} = \frac{3}{2}\left(1 + \mu_{\rm eff,p|s} \frac{\tilde{\mu}_{\rm p|s}(\omega)}{\mu_{\rm p|s}} \right)^{-1}.
    \label{eq:k2cmplx}
\end{eqnarray}
We adopt the Andrade viscoelastic model \citep{Andrade1910}, in which the complex shear modulus is given by:
\begin{eqnarray}
    \tilde{\mu}_{\rm p|s} {( \omega )} = \frac{1}{\tilde{J}_{\rm p|s} {( \omega )}},
    \label{eq:mucmplx}
\end{eqnarray}
where $\tilde{J}_{\rm p|s} {( \omega )}$ is the creep function, and $\omega$ is the angular velocity of the forcing cycle.

The creep functions, $\tilde{J}_\mathrm{p|s} {( \omega )}$, are given by \citep[e.g.,][]{Efroimsky2012a,Efroimsky2012CeMDA.112..283E}
\begin{eqnarray}
    \tilde{J}_\mathrm{p|s} {( \omega )} = {\left[ \frac{1}{\mu_\mathrm{p|s}} + \frac{1}{\mu_\mathrm{p|s} {\left( \tau_{\rm A} \omega \right)}^{\alpha}} \cos{\left( \frac{\alpha \pi}{2} \right)} \Gamma {\left( \alpha + 1 \right)} \right]} - i {\left[ \frac{1}{\eta_\mathrm{p|s} \omega} + \frac{1}{\mu_\mathrm{p|s} {\left( \tau_{\rm A} \omega \right)}^{\alpha}} \sin{\left( \frac{\alpha \pi}{2} \right)} \Gamma {\left( \alpha + 1 \right)} \right]},
    \label{eq:creep}
\end{eqnarray}
where $\eta_\mathrm{p|s}$ and $\eta_\mathrm{p|s}$ are the primary's or the satellite's efffective Andrade-Maxwell viscosity and shear rigidity, while  $\alpha = 0.33$ is the Andrade exponent of ice \citep[e.g.,][]{Rambaux+2010}.
The relaxation time of the Andrade model is given by $\tau_{\rm A} = \eta_\mathrm{p|s} / \mu_\mathrm{p|s}$.
Using Equations~(\ref{eq:k2cmplx})-(\ref{eq:creep}) the imaginary part of the complex Love number at a given $\omega$ frequency is
\begin{equation}
    \mathrm{Im}[k_\mathrm{2,p|s}(\tau_{\rm A} \omega)] = -\frac{3}{4}\frac{\mu_\mathrm{eff_\mathrm{p|s}} 
   \left((\tau_{\rm A} \omega) ^{\alpha }+\beta  \tau_{\rm A} \omega \right)}{ (\beta +\gamma  (\mu_\mathrm{eff,p|s}+1) \tau_{\rm A} \omega
   )+(\tau_{\rm A} \omega) ^{\alpha -1} \left((\mu_\mathrm{eff,p|s}+1)^2 (\tau_{\rm A} \omega
   )^2+1\right)+(\tau_{\rm A} \omega)^{1-\alpha} \left(\beta ^2+\gamma ^2\right)},
   \label{eq:Imk2}
\end{equation}
where
\begin{equation}
    \beta = \sin\left(\frac{\alpha \pi}{3}\right)\Gamma(\alpha+1),\,\,\,\,\,
    \gamma = \cos\left(\frac{\alpha \pi}{3}\right)\Gamma(\alpha+1).
\end{equation}
Since we do not model the thermal evolution of the bodies, we assume, for simplicity, that the relaxation time is temperature-independent (see, e.g., \citealp{Goldsby+2001}). 
Under this assumption, the viscosity is taken as constant, i.e., $\eta_{\mathrm{p|s}} = \eta_{\mathrm{ref}}$, where $\eta_{\mathrm{ref}}$ is calibrated to match the observed orbital distance of the satellite (see Section~\ref{sec:orbdistfit}).

\section{Computing Eccentricity Function}
\label{apx:E2k}

To accurately compute the tidal evolution of a planet–satellite system with large initial eccentricities ($e > 0.4$), it is necessary to replace the traditional series expansion of $E_{2,k}(e)$ based on Cayley’s tables with a more robust and accurate method. 
As shown by \citet{Ferraz-Mello2013CeMDA.116..109F}, the eccentricity function used in Equations~(\ref{eq:dadt})-(\ref{eq:dOmegasdt}) in CPL framework and Equation~(\ref{eq:BEA}) in ACT framework can be expressed as
\begin{equation}
    E_\mathrm{2,0,k}(e)=\frac{1}{2\pi\sqrt{1-e^2}}\int_0^{2\pi}\frac{1+e\cos(\nu)}{1-e^2}\cos(2\nu-(2+k)l)d\nu, 
    \label{eq:E2k}
\end{equation}
where $\nu$ and $l$ are the true and mean anomaly, respectively.
Following \citet{Smart1960ceme.book.....S}, the mean anomaly $l$ can be expressed as a series expansion in terms of the true anomaly $\nu$:
\begin{equation}
    l = \nu + 2 \sum^{m}_{j=1} (-1)^j \left( \frac{1}{j} + \sqrt{1 - e^2} \right) \gamma^j \sin(j\nu),
    \label{eq:mean-ano}
\end{equation}
where
\begin{equation}
    \gamma = \frac{1 - \sqrt{1 - e^2}}{e}.
    \label{eq:beta}
\end{equation}
For a given pair of values $(k,\,e)$, Equation~(\ref{eq:E2k}) is evaluated numerically using the trapezoidal rule to approximate the integral.

\begin{figure}
    \centering
    \includegraphics[width=0.45\columnwidth]{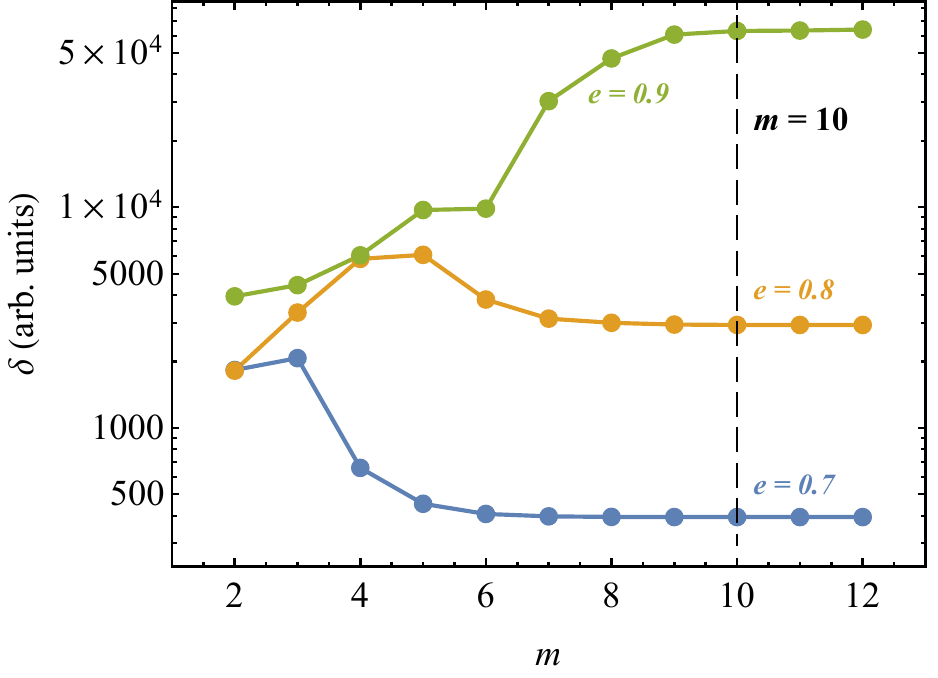}
    \includegraphics[width=0.435\columnwidth]{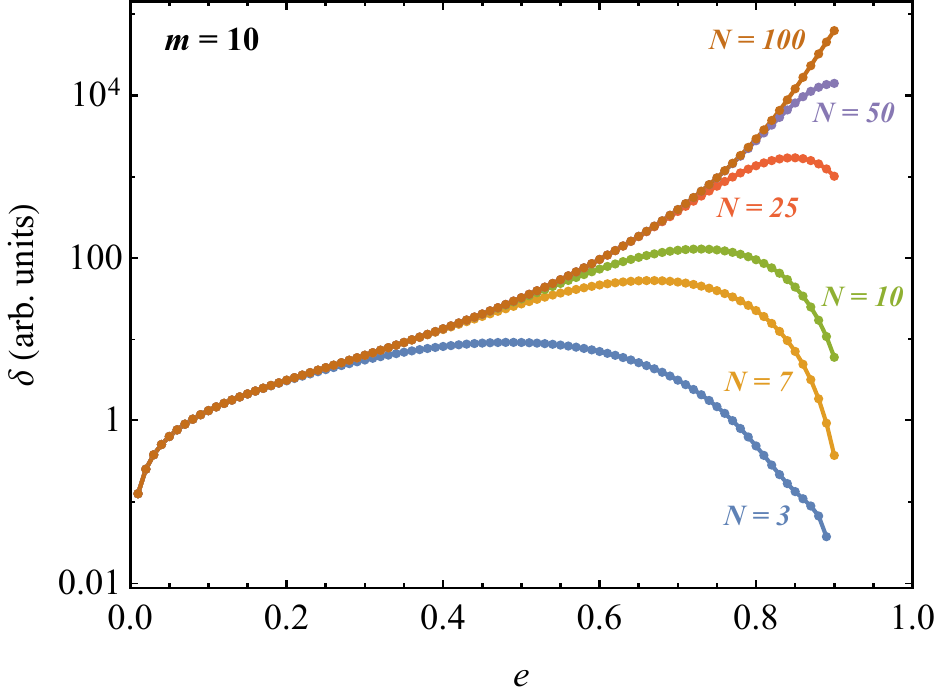}
    \caption{Left panel: Convergence of the eccentricity-dependent term Equation~(\ref{eq:delta}) as a function of the expansion order $m$ in the mean anomaly series.
    Results are shown for three different eccentricity values (color-coded), assuming a fixed summation limit of $N=100$.
    All curves exhibit convergence for $m \gtrsim 10$, justifying the choice of $m=10$ in our calculations.
    Right panel: Dependence of the eccentricity term Equation~(\ref{eq:delta}) on the eccentricity $e$, computed using $m=10$ and various summation limits $N$ (color-coded).
    As seen, models with $N < 100$ increasingly diverge from the $N=100$ reference solution at high eccentricities, indicating that high-order summation is essential for accurate modeling at $e \gtrsim 0.9$.
    Lower $N$ values lead to premature deviations at lower eccentricities.}
     \label{fig:delta-m-e}
\end{figure}

To determine the appropriate order of expansion $m$ in Equation~(\ref{eq:mean-ano}), we evaluate the leading eccentricity-dependent term from the CPL framework, as described by Equation~(\ref{eq:dedt}). This term is given by:
\begin{equation}
    \delta = \frac{1}{e} \sum_{k=-N}^{N} \left[ \left( 2\sqrt{1 - e^2} - (2 - k)(1 - e^2) \right) E^2_{2,0,k}(e) \, \mathrm{sgn}(2\Omega_\mathrm{p} - (2 - k)n) \right],
    \label{eq:delta}
\end{equation}
where $\Omega_\mathrm{p}$ is the rotation rate of the primary and $n$ is the satellite's mean motion. This expression allows us to assess the convergence of the eccentricity series expansion by examining the sensitivity of $\delta$ to the truncation order $m$.
Assuming the primary is in synchronous stationary rotation, the sign function simplifies to $\mathrm{sgn}(2\Omega_\mathrm{p} - (2 - k)n) = \mathrm{sgn}(k)$.
The left panel of Figure~\ref{fig:delta-m-e} displays the magnitude of $\delta$ as a function of the expansion order $m$, assuming $N = 100$, for three representative high eccentricity values. 
As shown, convergence requires a relatively high expansion order, particularly at large eccentricities ($e \geq 0.9$), where $m \gtrsim 10$ is necessary to obtain stable results.
Based on these findings, we adopt $m = 10$ in our calculations to ensure sufficient accuracy across the full eccentricity range explored in this study.

It is important to note that the summations in Equation~(\ref{eq:delta}) are computationally intensive, particularly for large values of $N$. 
The right panel of Figure~\ref{fig:delta-m-e} illustrates the dependence of $\delta$ on eccentricity $e$ for various values of $N$. 
As shown, accurate computation of $\delta$ at high eccentricities ($e \gtrsim 0.9$) requires a summation range of at least $N = 100$. 
In contrast, for moderate eccentricities, lower values of $N$ are sufficient: $N = 10$ ensures accuracy for $e \leq 0.5$, and $N = 3$ is adequate for $e \leq 0.2$.
To optimize computational efficiency while maintaining accuracy, we adopt an adaptive approach in which the value of $N$ is selected based on the current eccentricity of the satellite.

\begin{figure}
    \centering
    \includegraphics[width=0.6\columnwidth]{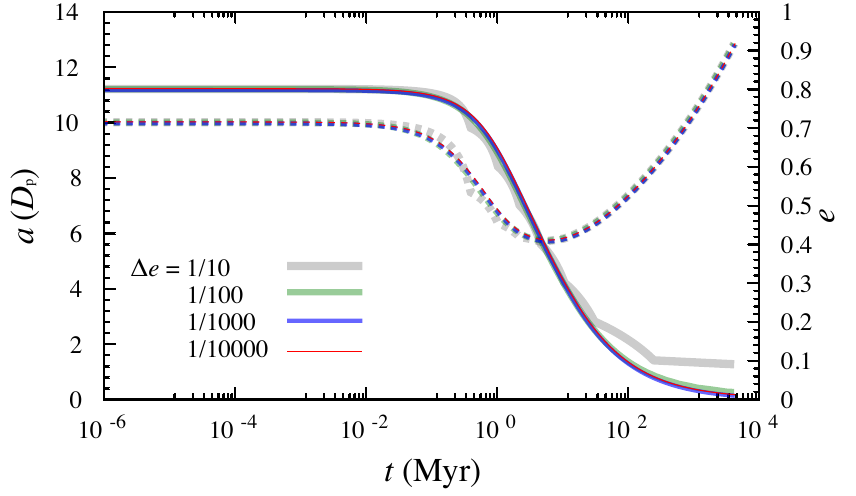}
    \caption{4.5 Gyr tidal evolution of the Quaoar–Weywot system computed using the numerical approximation of the eccentricity function $E_\mathrm{2,0,k}(e)$ via the precomputed matrix ${\bf E}_{2,0}[k,e]$.
    Different colors represent various resolutions in eccentricity used to construct ${\bf E}_{2,0}[k,e]$.
    Solid lines show the evolution of Weywot’s eccentricity, while dashed lines indicate the evolution of its semimajor axis.
    All models assume an initial semimajor axis of $a_\mathrm{ini} = 10~D_\mathrm{p}$, initial eccentricity $e_\mathrm{ini} = 0.8$, and initial spin period of Quaoar $P_\mathrm{p,ini} = 5$~hr.
    Weywot is assumed to be initially in synchronous rotation.}
    \label{fig:e_convtest}
\end{figure}

To compute the tidal evolution using either the CPL framework described by Equations~(\ref{eq:dadt})--(\ref{eq:dOmegapdt}) or the ACT framework given by Equations~(\ref{eq:BEdadt})--(\ref{eq:BEdOmegapdt}), it is necessary to evaluate the eccentricity functions $E_{2,0,k}(e)$ for a given eccentricity $e$. 
Since these frameworks require summations over a wide range of harmonic indices ($-N \leq k \leq N$), it is computationally efficient to precompute the eccentricity functions over the domain $-200 \leq k \leq 200$ and $0 < e < 1$. This yields a two-dimensional lookup table or matrix, denoted as ${\bf E}_{2,0}[k,e]$.
During the simulations, the required value of $E_{2,0,k}(e)$ can then be efficiently approximated by accessing the corresponding element of ${\bf E}_{2,0}[k,e]$ through interpolation. 
Based on our numerical convergence tests (Figure~\ref{fig:e_convtest}), discretizing the eccentricity interval $0 \leq e \leq 1$ into 10,000 uniformly spaced points ensures sufficient accuracy for the purposes of this study.
For the ACT framework, a similar methodology was employed to construct the look-up table ${\bf E}_{2,1}[k,e]$, representing the eccentricity function corresponding to $E_{2,1,k}(e)$ given by Equation~(\ref{eq:E2pk}).

\end{document}